\documentclass[preprint2]{aastex}
\usepackage{emulateapj5}
\usepackage{onecolfloat5}

\newcommand{\beq}   {\begin{equation}}
\newcommand{\eeq}   {\end{equation}}
\newcommand{\mua}   {\mu_{\alpha}}
\newcommand{\mud}   {\mu_{\delta}}

\newcommand{\muvs}  {\mbox{\boldmath\scriptsize $\mu$}}

\newcommand{\xxs}   {\mbox{\bf\scriptsize X}}

\newcommand{\vvs}   {\mbox{\bf\scriptsize V}}

\newcommand{\zhat}  {\hat{\mbox{\bf z}}}

\newcommand{\vvperp} {\mbox{\bf\scriptsize V}_{\perp}}

\newcommand{\kms}   {km~s$^{-1}$}
\newcommand{\aam}   {\altaffilmark}
\newcommand{\aat}   {\altaffiltext}

\shortauthors{Vlemmings, Cordes \& Chatterjee}
\shorttitle{Separated at birth: B2020+28 and B2021+51}
\slugcomment{For the Astrophysical Journal}

\begin{document}
\twocolumn[
\title{Separated at Birth: The origin of the pulsars B2020+28 and
  B2021+51 in the Cygnus Superbubble} 
\author{ W. H. T. Vlemmings\aam{1}, J. M. Cordes\aam{1}, S. Chatterjee\aam{2}}

\begin{abstract}
 High precision astrometric data have enabled us to determine the
trajectories through the Galactic potential for a growing number of
pulsars. This has resulted in the discovery of a pulsar pair (B2020+28
and B2021+51) that has a common origin in the Cygnus Superbubble or in
one of its related OB associations at an epoch which is comparable
with the spin-down ages of the pulsars. Analysis of the Galactic
orbits indicates that the progenitors of the pulsars had similar
masses and were in a binary system, which was disrupted after the
second supernova explosion. The implied pulsar birth velocities are
consistent with the high velocities of neutron stars in general. The
initial spin period of the pulsar that was formed in the second
supernova explosion was $\sim 200$~ms. A further increase in
astrometric accuracy will allow us to more tightly constrain the birth
velocities and the kick velocities that were imparted by the two
respective supernova explosions. Two additional pulsars in the sample
of 24 with parallax measurements may also have originated in the Cygnus
Superbubble.
\end{abstract}
]

\aat{1}{Department of Astronomy, Cornell University, Ithaca, NY 14853;
wouter@astro.cornell.edu.}
\aat{2}{Jansky Fellow, National Radio Astronomy Observatory, P.O. Box
  O, Socorro, NM 87801.}

\section{Introduction}

 The velocity distribution of pulsars has long ago been shown to
extend to much larger velocities then those of their progenitors
\citep{GO70}. Several possible scenarios have been suggested as causes
for high pulsar velocities. These include asymmetry in the birth
supernovae (SN) of pulsars \citep{S70, DC87} and the disruption of
binaries by mass-loss from the SN explosion \citep{B61,GGO70}.
Considerable evidence exists that both effects take place.  Exactly
how an asymmetry in the core collapse of a SN explosion is converted
to a kick imparted on the newly formed neutron star (NS) is still
unclear; possibilities include hydrodynamic or convective
instabilities \citep{BH96, JM96, LG00} and mechanisms such as
asymmetric neutrino emission in the presence of strong magnetic
fields \citep{AL99}.

Binarity is particularly common among the early type O and B stars
that are expected to produce Type II SN and subsequently NS;
approximately 70\% of these stars exist in binary systems
\citep{B67}. In about 50\% of the binaries, both stars are
massive enough to become SN; most of these binaries are disrupted when
the first SN occurs, while only a small percentage evolves into NS-NS
binaries \citep{BB98}. Until now 6 of such NS-NS binaries have been
detected \citep{burg03}.

While most massive star binaries are disrupted at either the first or
the second SN, many individual pulsars originate from such
binaries. Identification of these pulsars will allow us to put
constraints on kick velocities and the amount of asymmetry in the SN
explosions.

Previously, several pairs pulsars where hypothesized to come from the
same disrupted binary based on their positional information alone
\citep{GGO70, W79}.  Recently, high precision VLBI astrometry has
produced accurate measurements of parallaxes and proper motions of
about two dozen nearby pulsars, which allows much more accurate tests
for pulsar pair identifications. Using the astrometric data, we have
analyzed possible trajectories for the pulsars that take into account
the Galactic gravitational potential and plausible values of the
unknown radial velocities.  We find that two of the pulsars (B2020+28
and B2021+51) have Galactic orbits that cross at an epoch consistent
with their spin-down ages, which makes them prime candidates for
pulsars that have been ejected from the same binary system. In this paper we
first discuss the identification of the pulsar pairs and describe the
Galactic potential used to calculate the pulsar orbits
(\S~\ref{sec2}).  The results of the orbit calculations are presented
in \S~\ref{sec3}. The birth location of the pulsars and several
possible birth scenarios are discussed in \S~\ref{sec4}, followed by
our conclusions in \S~\ref{concl}.

\section{Identification of pulsar pairs}
\label{sec2}

 The sample of pulsars with high precision astrometric data to date
consists of 24 objects, six of which are recycled pulsars. The
complete sample is further discussed in Chatterjee et al. (2004, in
preparation). First, the trajectories of the 24 pulsars were
determined for 3 different radial velocities ($v_{\rm r}=-300, 0$ and
$+300$~\kms) in the Galactic potential which is discussed below
(\S~\ref{galpot}). We then further examined the Galactic orbits of
those pulsar pairs that cross on the plane of the sky at comparable
distances and at a similar epoch. From this analysis we find that
B2020+28 and B2021+51 are likely candidates for a pulsar pair with a
common origin in the Cygnus Superbubble (CSB). Two other pulsars
(B1857$-$26 and B2016+28) in our sample may also have originated from
this region.

\subsection{B2020+28 and B2021+51 and their Galactic Orbits}

\label{pdf}

 The astrometric properties of B2020+28 and B2021+51, which are listed
in Table~\ref{Table:astrom}, imply transverse velocities of
$307$~\kms\ and $120$~\kms\ for B2020+28 and B2021+51 respectively.
The measured period ($P$), period derivative ({$\dot{P}$}) and
dispersion measure ($DM$) are give in Table~\ref{Table:pulsar}, along
with the spin-down age, $\tau_{\rm sd} = P/{2\dot{P}}$ , and the
magnetic field strength $B = 3.2\times10^{19}\sqrt{P\dot{P}}$, both
for a braking index $n=3$. Neither of the pulsars appears to have been
recycled through accretion in a binary system, as the values for $B$
are consistent with canonical, non-recycled values.

To determine the minimum distance between the pulsar pair at earlier
epochs, we performed Monte Carlo simulations of the Galactic pulsar
orbits taking into account the astrometric uncertainties and the
unknown radial velocities. We assumed a Gaussian probability
distribution for the proper motion and parallax measurements, $\mua$,
$\mud$ and $\pi$, with standard deviations, $\sigma_{\mu_\alpha}$,
$\sigma_{\mu_\delta}$ and $\sigma_\pi$, respectively. The present day
absolute positional uncertainty, estimated at 15 mas, corresponds to
only $2\times10^{-4}$~pc at 2 kpc distance and is negligible, as the
proper motion errors already correspond to a positional uncertainty of
$\approx 5$~pc when extrapolated backward in time by 1~Myr. The
distribution of the radial velocity was determined assuming a two
component pulsar birth velocity distribution ($f_{\vvs_0}$) as
determined by Chatterjee et al.~(2004, in preparation) and consistent
with the work of \citet{CC98} and \citet{ACC02}. The birth velocity
PDF was corrected to first order for the propagation through the
Galactic potential for a time corresponding to the age $\tau$ of the
pulsar. This gives the current velocity PDF:

\beq 
f_{\vvs}(\vvperp,v_{r}) = f_{\vvs}(\vvs) = f_{\vvs_0} (\vvs + \tau
\ddot{z}\zhat) 
\eeq 
where we have estimated the true age to be equal to the spin-down age,
$\tau=\tau_{\rm sd}$, and where $\ddot{z}$ is the acceleration due to
the Galactic potential in the $z-$direction perpendicular to the
Galactic plane. Note that discrepancy of the true age from the
spin-down age are not significant in our analysis, because we restrict
our study to young objects and thus the correction due to the
acceleration is small. Then, given the transverse velocities, the
conditional PDF for the radial velocity is:

\beq
f_{v_r}(v_r) = \int d\vvperp~f_{\vvs}(\vvperp,v_{r}) f_{\vvperp}(\vvperp)
\label{func_0}
\eeq 
where $f_{\vvperp}(\vvperp)$ includes the PDFs of the measurement
errors in proper motion and parallax. This conditional PDF was then
used to randomly select the pulsar radial velocity in each Monte Carlo
trial. The conditional PDFs for B2020+28 and B2021+51 are shown in
Fig.~\ref{Fig:radpdf}.

\begin{figure*}[ht]
\epsscale{1.6} 
\plotone{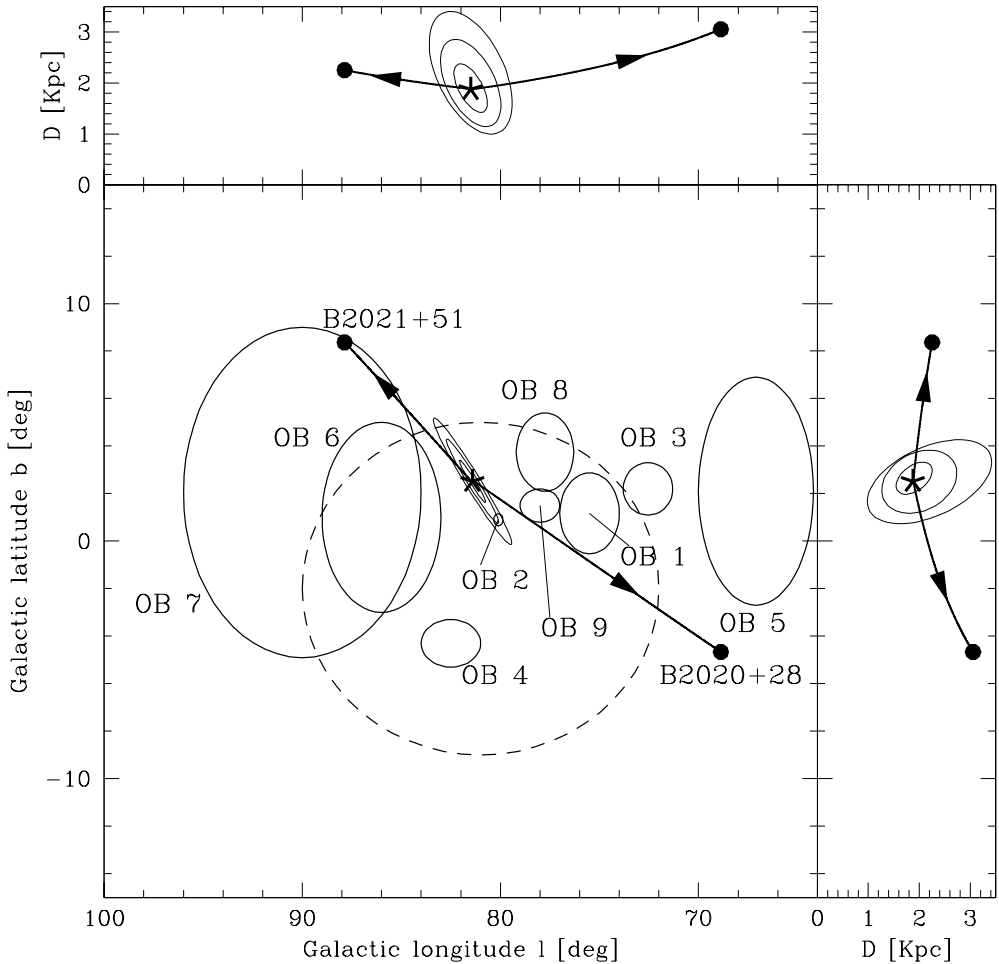}
\caption[The pulsar path through the Cygnus superbubble]{ The
3-dimensional pulsar motion through the Galactic potential is shown
for one of the pulsar orbit solutions that yields a minimum separation
$< 10$~pc (these particular Galactic orbits cross within 4 pc). The
dashed circle represents the Cygnus superbubble, while the labeled
solid ellipses are the Cygnus OB associations with positions and
extents as tabulated by \citet{UFR+01}. The extent of OB 2 is unknown
and only the center of the association is indicated. The thick solid
lines indicate the pulsar paths, with the origin denoted by the
starred symbol and the arrows pointing in the direction of motion. The
current positions are indicated by the solid dots. The elliptical
contours around the pulsars' origin in these panels indicate the 1, 2
and 3$\sigma$ levels of the likelihood solution for the birth location
for Galactic orbit solutions that reach a minimum separation $<
10$~pc.}
\label{Fig:cygnuspath}
\end{figure*}

\begin{deluxetable}{lcccccc}
\tablecolumns{7}
\tablewidth{0pc} 
\tablecaption{Pulsar Astrometric data\label{Table:astrom}}
\tablehead{ 
\colhead{Pulsar} & \colhead{$\alpha_{\rm J2000.0}$} &
\colhead{$\delta_{\rm J2000.0}$} & \colhead{$\mua$} & 
\colhead{$\mud$} & \colhead{$\pi$} & \colhead{$D$}\\
\colhead{} & \colhead{ } & \colhead{ } & \colhead{(mas yr$^{-1}$)} &
\colhead{(mas yr$^{-1}$)} & \colhead{(mas)} & \colhead{(kpc)}
}
\startdata
B2020+28  & 20~22~37.0718 & 28~54~23.0300 & $-4.38\pm0.53$ &
$-23.59\pm0.26$ & $0.37\pm0.12$ & $2.7^{+1.3}_{-0.7}$\\ 
B2021+51  & 20~22~49.8655 & 51~54~50.3881 & $-5.23\pm0.17$ &
~$11.54\pm0.28$ & $0.50\pm0.07$ & $2.0^{+0.3}_{-0.2}$\\ 
\enddata 
\tablecomments{The data are from \citet{BBGT02}. The positions are at
  the epoch 2000.0.} 
\end{deluxetable} 

The Galactic orbits of the pulsars cross within $10$~pc
for $\approx 0.15\%$ of our Monte Carlo simulations.  The resulting
Galactic orbits of B2020+28 and B2021+51 are shown in
Fig.~\ref{Fig:cygnuspath} and correspond to co-location $\sim~1.9$~Myr
ago. The figure indicates the pulsar trajectories as seen from an
observer at rest in the Solar reference frame and includes the
Galactocentric motion of the Sun. Only one of the Galactic orbit pairs
that results in a minimum separation $<10$~pc between the two pulsars
is shown. The contours around the common origin position (starred
symbol) indicate the $1,2$ and $3\sigma$ likelihood contours of the
birth location for Galactic orbit solutions which approach 10~pc. The
pulsars are seen to traverse the Cygnus Superbubble region (dashed
ellipse), which contains several OB associations (see
\S~\ref{cygnus}). While there are multiple orbit solutions that yield
minimum separations $< 10$~pc and such co-locations can occur by chance
for unrelated pairs, we demonstrate below that the probability of
co-location is in this case too large to have been caused by chance.
Also the choice of Galactic potential does not significantly alter the
results, as the Galactic orbit integration is only
$\approx~2$~Myr. Even when no Galactic potential is used, a
significant number of pulsar orbits reach a minimum separation
$<10$~pc.

\begin{deluxetable}{lccccc}
\tablecolumns{6}
\tablewidth{0pc} 
\tablecaption{Pulsar Characteristics\label{Table:pulsar}}
\tablehead{ 
\colhead{Pulsar} & \colhead{$P$} & \colhead{$\dot{P}$} & \colhead{$DM$} &
\colhead{$\tau_{\rm sd}$} & 
\colhead{$B$} \\
\colhead{} & \colhead{(s)} & \colhead{($\times 10^{-15}$~s~s$^{-1}$)} & \colhead{(cm$^{-3}$~pc)} & \colhead{(Myr)} &
\colhead{($\times 10^{12}$ G)}
}
\startdata
B2020+28  & $0.343$ & $1.89$ & $24.62$ & $2.88$ & $0.81$ \\
B2021+51  & $0.529$ & $3.07$ & $22.58$ & $2.75$ & $1.29$ \\
\enddata 
\end{deluxetable} 

\subsection{Galactic Pulsar Orbits}
\label{galpot}

The pulsar orbits were traced back in time through a three component
St\"ackel potential, which satisfies the most recent estimates of the
Milky Way parameters \citep[][hereafter FD03]{FD03}. The three
components include a thin disk, a thick disk and a halo component. As
shown in FD03, a bulge component did not have to be included
explicitly, as the resulting three component potential turns out to
simulate an effective bulge. In FD03 several combinations of component
parameters are shown to satisfy the Milky Way parameters. In our
potential we chose the axis ratios $\epsilon = 75.0, 1.5$ and $1.02$,
and relative contributions $k = 0.13, 0.01$ and $1.0$ for the thin
disk, thick disk and halo respectively. The scaling parameters were
then adjusted so that our potential predicts the Oort constants $A =
14.19$ \kms~kpc$^{-1}$, $B = -11.69$ \kms~kpc$^{-1}$ and a circular
velocity $v_{\rm circ} = 219.7$ \kms\ at $R_\odot =
8.5$~kpc. Additionally, our potential has a mass density in the solar
neighborhood of $\rho_\odot \approx 0.09$~M$_\odot$~pc$^{-3}$.  These
values agree with those obtained in an analysis of Hipparcos data by
\citet{FW97}.  The local mass density is also in agreement with
$0.06$~M$_\odot$~pc$^{-3} < \rho_\odot < 0.12$~M$_\odot$~pc$^{-3}$
determined from Hipparcos observations \citep{CCB98, HF00, SBS03}. We
assume that perturbations of the Galactic orbit due to small scale
structure in the potential are negligible, as our integration times
are less then a few Myr.

From the astrometric data we determine the position vector in the
Galactic reference frame and the velocity $\vvperp$ in the plane of the
sky. The pulsar velocity $\vvs_{\rm psr}$ is then calculated with a
radial velocity taken from the conditional PDF for radial velocity
$v_{\rm r}$ as described above. After correcting the pulsar velocity
$\vvs_{\rm psr}$ for solar motion with respect to the
local standard of rest (LSR) $\vvs_{\rm sun}$, differential Galactic
rotation $\vvs_{\rm DGR}$, and the velocity of the LSR $\vvs_{\rm LSR}$, the
pulsar velocity in the Galactic reference frame is:

\beq
\vvs'_{\rm psr} = \vvs_{\rm psr} - \vvs_{\rm sun} - \vvs_{\rm DGR} - \vvs_{\rm LSR}. 
\eeq
This velocity is then used to retrace the pulsar orbit in the Galactic
potential with a fourth order Runge-Kutta numerical integration method
and fixed time steps of 1000 years.

\subsection{Identification of pulsar pairs}
\label{positive}
\label{chance}

 A positive pair identification is assumed when simulations of the
Galactic orbits produce a significant number of solutions where the
minimum orbit separation is $< 10$~pc at a realistic epoch. At the
distance to the pulsars discussed here as determined by the parallax
measurements, 10~pc correspond to only $\approx 5$~\kms\ uncertainty
in space velocity during $\approx 2$~Myr. As was shown in
\citet{HdBdZ01}, trajectories that cross within a few pc are only a
small fraction of all the orbits consistent with the measurement
errors. Here we numerically estimate the probability to detect a
pulsar pair within $10$~pc of each other when taking into account the
measurement uncertainties on the astrometric parameters.

The conditional PDF for the pulsar separation $\Delta_d$ at time $t$ is given by:
\beq
\label{func_1}
f_{\Delta_{d}} (\Delta_{d}|v_{r_1},v_{r_2},\muvs_{1},\muvs_{2},\pi_{1},\pi_{2},t) = \delta(\Delta_{d} - |\xxs_{t_2} - \xxs_{t_1}|)
\eeq
Here $\muvs_{i}$ and $\pi_{i}$ are the astrometric data for the two pulsars and $v_{r_{i}}$ is the radial velocity. The separation between the pulsars at time $t$ is:
\beq
\Delta_d = |\xxs_{t_2} - \xxs_{t_1}|,
\eeq
with
\beq
\xxs_{t} = G(\xxs,\vvs,t).
\eeq
The function $G$ describes the trajectory of the pulsar and
incorporates the effects of the Galactic potential. In the absence of
the Galactic potential $G=\xxs-\vvs t$. Eq.~\ref{func_1} is integrated
over the astrometric uncertainty PDFs as well as over the radial
velocity PDF given by Eq.~\ref{func_0}. To obtain the probability for
a minimum separation $< 10$~pc between the pulsars we integrate
further over the separation $\Delta_{d} < 10$~pc and over the times $t
= t_{\rm min}$ at which this minimum separation occurs, \beq
\label{func_2}
P_{c}(\tau_0) = \int_{t_{\rm min}} dt \int_{\Delta_{d} < 10~{\rm pc}} d \Delta_{d} f_{d_{t}}.  
\eeq

\begin{figure*}[tbf]
\epsscale{1.7} 
\plotone{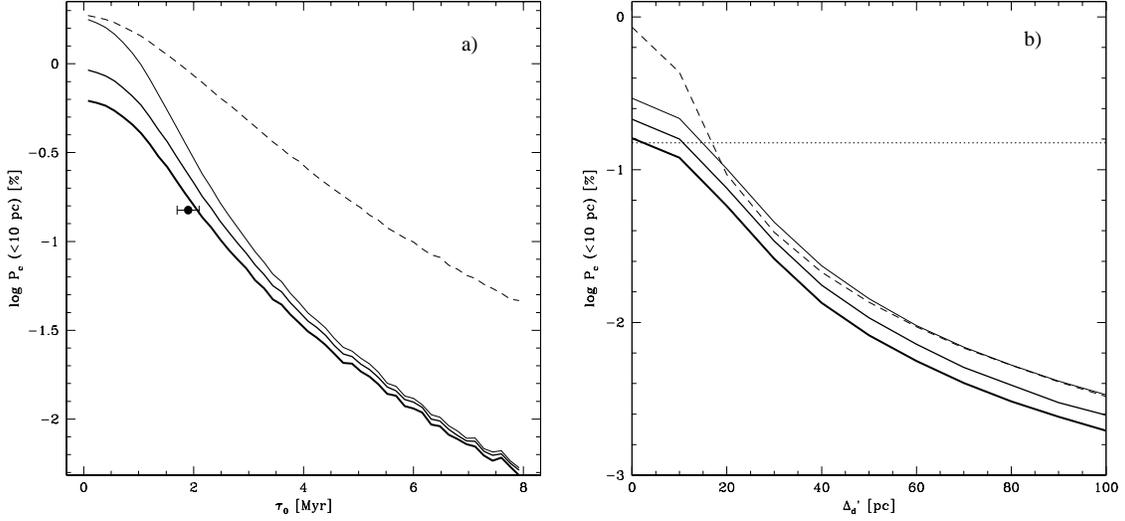}
\caption[detection probabilities]{The probability function $P_c$ for
finding pulsar orbits with a minimum separation $<10$~pc. Left panel
(a): as a function of the time before the present $\tau_0$ at which
the (simulated) pulsars are known to be co-located. For larger
$\tau_0$, the present day proper motions are extrapolated further back
in time; the growing errors on the predicted pulsar positions cause
the probability of finding pulsar with a minimum separation $<10$~pc
to decrease. The thick solid line indicates $P_c$ for a pulsar pair
with astrometric uncertainties similar to those of B2020+28 and
B2021+51. The thinner lines correspond to 30\% and 60\% improvements
on the parallax uncertainties, while the dashed line corresponds to a
pulsar pair with uncertainties that are 30\% of the $\mua$, $\mud$ and
$\pi$ uncertainties of B2020+28 and B2021+51. For $\tau_0 > 4$~Myr the
curves show some numerical noise. The point indicates the percentage
of Galactic orbit solutions with a minimum separation $<10$~pc for our
observed pulsar pair with the estimated $1\sigma$ errors on the
determined $\tau_0$. Right panel (b): as a function of {\it known}
minimum separation at time $\tau_0 = 2~$Myr. The assumed astrometric
uncertainties for the four curves are similar to those in the left
plot. The horizontal dotted line corresponds to $P_c$ determined for
B2020+28 and B2021+51.}
\label{Fig_prob}
\end{figure*}

As $t_{\rm min}$ depends on the time $\tau_0$ at which the actual
minimum separation occurs, the resulting probability function in
Eq.~\ref{func_2} is a function of $\tau_0$ that reflects 
errors that increase as the astrometric properties of the pulsars are
projected back in time. Using astrometric errors similar to those
determined for B2020+28 and B2021+51, Eq.~\ref{func_2} is evaluated
numerically for a pulsar pair that is known to coexist at $\tau_0$,
limiting the analysis to pulsar pairs with $\tau_0 < 10~$Myr. The
resulting distribution is shown in Fig.~\ref{Fig_prob}a. We find that
the probability of finding pulsar orbits that cross within $< 10$~pc
is $P_{c}(2~$Myr$) \approx 0.16\%$, for a pulsar pair that is known to
have a common origin at $\tau_0 = 2~$Myr and that has current day
astrometric errors similar to the measurements. This is remarkably
consistent with the value of $0.15\%$ found for B2020+28 and B2021+51.

If, instead of being produced in a binary, the pulsar pair crosses at
a known minimum distance $\Delta_{d}'$ at time $\tau_0$, the
probability of finding orbits within $\Delta_{d} < 10$~pc decreases
with $\Delta_{d}'$ as shown in Fig.~\ref{Fig_prob}b. Here we again
used astrometric errors similar to those of B2020+28 and B2021+51, for
a pulsar pair that has an actual minimum separation $\Delta_{d}'$ at
$\tau_0 = 2~$Myr. Thus, for $\Delta_{d}' = 0$, the value of
$P_{c}(\Delta_{d}'=0) = P_{c}(\tau_0=2~$Myr$) = 0.16\%$.  $P_c$ drops
sharply for increasing $\Delta_{d}'$, especially when the astrometric
errors in proper motion are decreased.  The effect of a reduction of
parallax uncertainties is less, as these are generally larger than the
several tens of parsec distance scale which was examined for
$\Delta_{d}'$.

\begin{figure*}[ht!]
\epsscale{1.8} 
\plotone{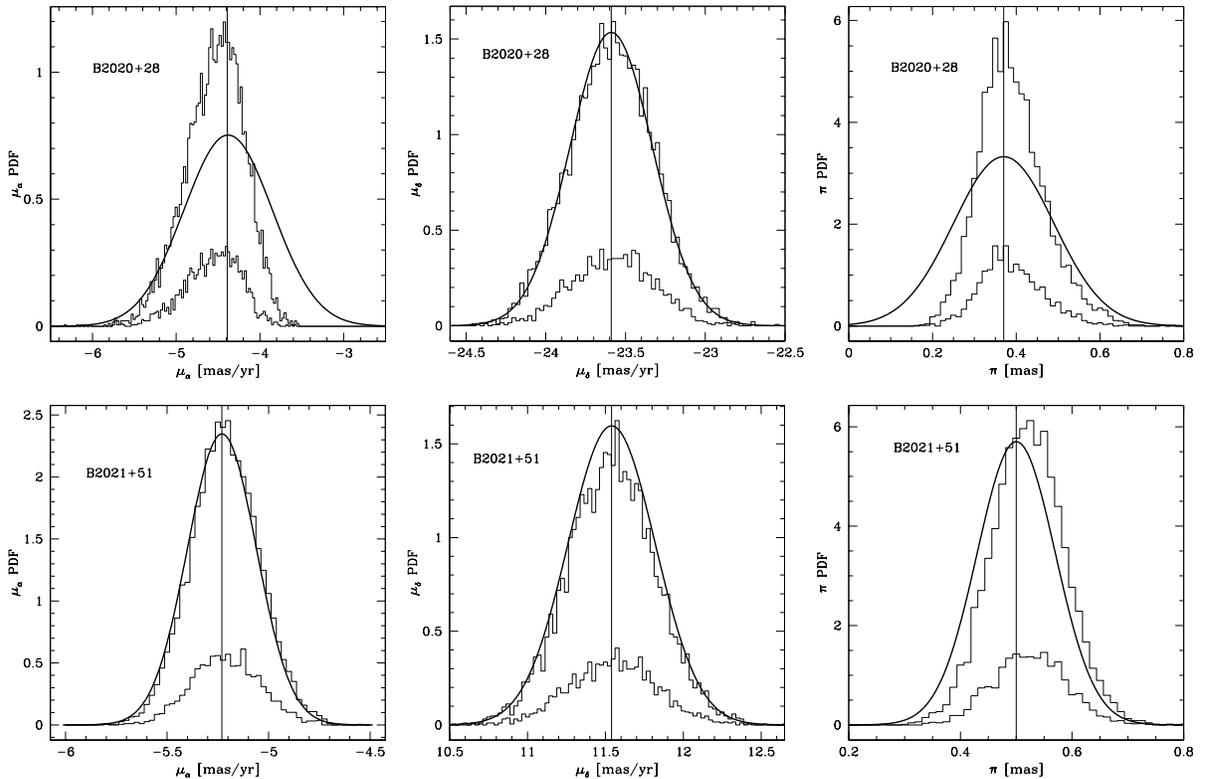}
\caption[astrometric pdfs]{ Conditional distributions of the
astrometric variables $\mua$ (left), $\mud$ (middle) and $\pi$ (right)
for the Galactic orbit solutions that yield minimum separations $<
10$~pc (thick histogram) and $< 5$~pc (thin histogram). The top panels
are for B2020+28 and the lower panels for B2021+51. The observed
values are indicated by the vertical solid lines and the thick solid
curve indicate the initial Gaussian PDFs corresponding to the
astrometric errors. Note that both histograms are normalized with the
same factor.}
\label{Fig:pdf}
\end{figure*}

To further investigate if the identification of the binary pulsar pair
is only a chance coincidence, we simulated 20 samples of nearby ($< 3$
kpc) pulsars using the birth velocity and scale height distribution as
determined by Chatterjee et al.~(2004). The samples consisted of 18
pulsars, matching the current sample of non-recycled objects with high
precision astrometry. We restricted our simulations to young ($<
25$~Myr), non-recycled objects. The pulsars were propagated in the
Galactic potential to determine their astrometric properties. The
radial velocity information was discarded and we determined the
closest approach for each of the 153 pulsar pairs in the samples by
propagating the pulsars back in time on a grid of possible radial
velocities ($|v_{\rm r}| < 800$ \kms).  We found that $\approx 80\%$
of our samples could produce one or more pulsar pairs with a minimum
separation $< 10$~pc. However, the identified pairs are heavily biased
toward pulsar ages $>10$~Myr. We do not identify any pair younger
than $5~$Myr. Those pairs that produce the most solutions in this
first, coarse, analysis, were then investigated further using the same
Monte Carlo method as described in \S~\ref{pdf}. We included
uncertainties on the proper motions and parallaxes that were similar
to the astrometric uncertainties for B2020+28 and B2021+51, and the
radial velocity was drawn from the actual velocity PDF determined for
each pulsar as shown in Eq.~\ref{func_0}. We found that the likelihood
of finding a minimum separation $< 10$~pc for these simulated pairs is
generally $< 10^{-4}\%$. This low probability rules out any spurious
identification as a real related pulsar pair for all of the pulsars in
our simulated samples.

 However, instead of being a disrupted binary, the pulsars might have
originated in one or two of the nearby OB associations, which have
enhanced SN rates. Therefore we simulated a large number of pulsars
coming from a spherical volume with a radius of approximately 0.8 kpc,
which is the volume that encompasses several of the OB associations
found in the Cygnus superbubble region (See \S~\ref{cygnus}). We
simulated 40 pulsars, corresponding to 780 possible pulsar pairs,
which is approximately 20\% of the number of pulsars estimated to have
formed in the CSB \citep{UFR+01}. In an analysis similar to
the one performed on our simulated Galactic pulsar samples, we find
that approximately 5\% of the pulsar pairs can be traced back to
within 10 pc for realistic radial velocities. However, in a further
analysis of those pairs, we find that the likelihood of a minimum
separation $<10$~pc is only $0.01\%$, much less than the
likelihood expected for a real pulsar pair. As seen in
Fig.~\ref{Fig_prob}b, this is consistent with an actual minimum
separation of $>40$~pc. Thus, we are confident that B2020+28 and
B2021+51 are indeed related and originate from the same disrupted
binary.

\section{Astrometric parameters and ages of B2020+28 and B2021+51}

\label{sec3}

\begin{figure}[t]
\epsscale{1.0} 
\plotone{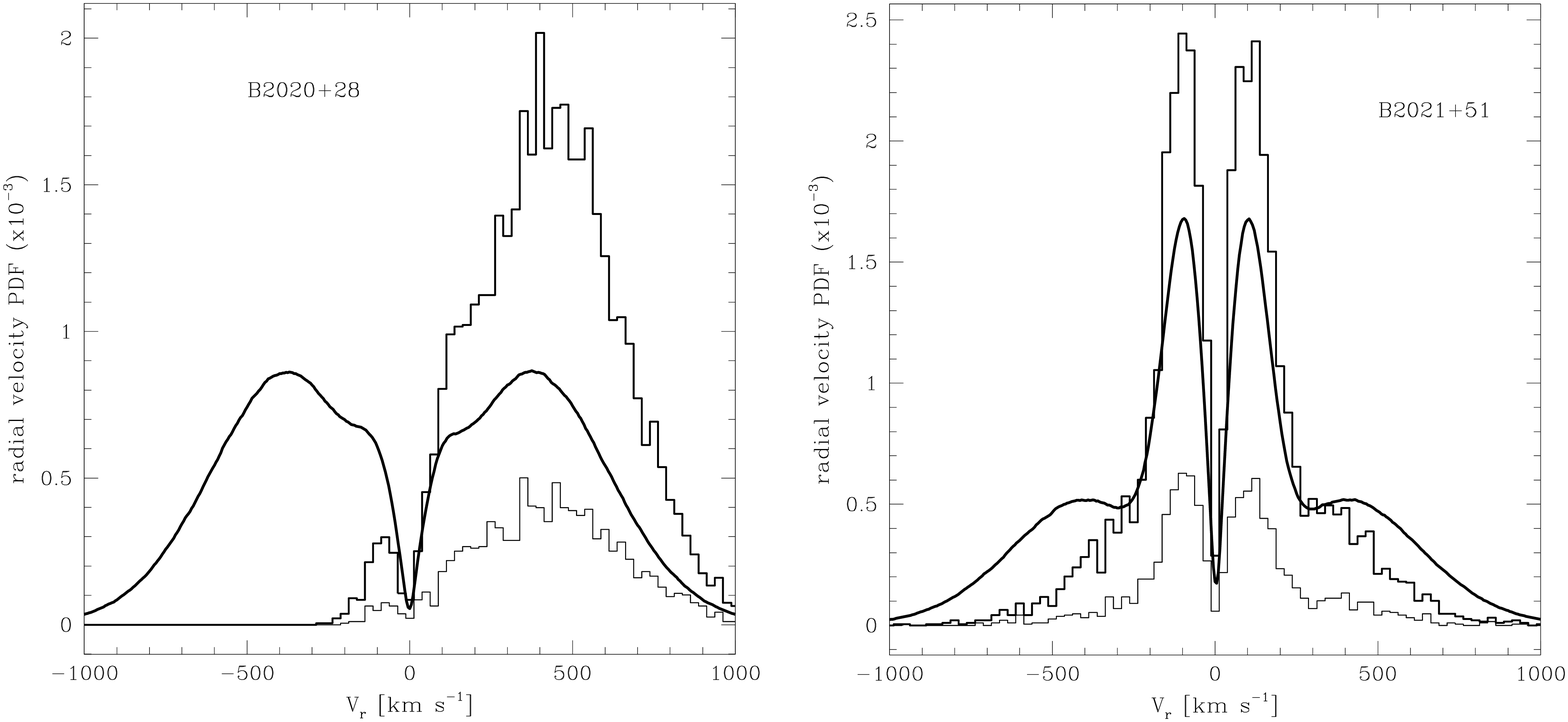}
\caption[radial velocity pdfs]{ Conditional distribution of the radial
velocities of B2020+28 (left) and B2021+51 (right) for the Galactic
orbit solutions yielding minimum separation $< 10$~pc (thick
histogram) and $< 5$~pc (thin histogram). The initial conditional PDF
for $v_{\rm r}$, as described in \S~\ref{pdf}, is indicated by the
thick solid curve. The normalization of both histogram is similar.}
\label{Fig:radpdf}
\end{figure}

\begin{figure}[htf]
\plotone{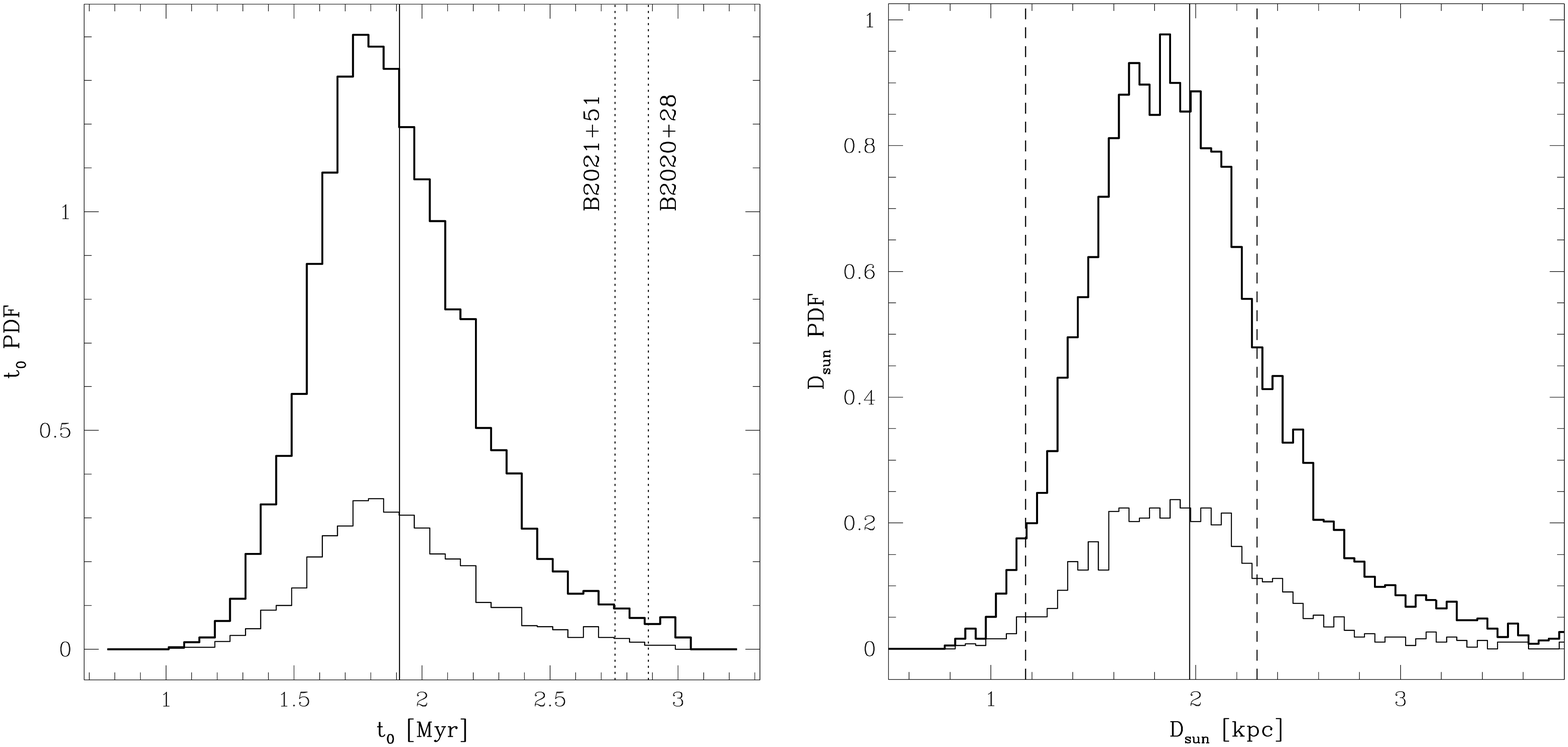}
\caption[t-dsun pdfs]{ Distribution of the times $\tau_0$ (left) and
distance to the sun $D_{\rm sun}$ (right) for the Galactic orbit
simulation results yielding minimum separations $< 10$~pc (thick
solid) and $< 5$~pc (thin solid) between B2020+28 and B2021+51. The
solid vertical lines denote the mean of the distributions. In the left
figure, the dotted vertical lines indicate the spin-down ages of the
pulsars. In the right figure, the short dashed vertical lines indicate
the estimated distance interval to the Cygnus superbubble and OB
associations. The normalization of both histograms is similar.}
\label{Fig:closepdf}
\end{figure}

The pulsar pair B2020+28 and B2021+51 was found to have a minimum
separation $< 10$~pc for $\approx 0.15\%$ of 3 million Monte Carlo
runs. Here we demonstrate that the values of $\mua$, $\mud$ and $\pi$
required for these orbits are typical values allowed by the PDF's for
those quantities, and not special values with low
probability. Figs.~\ref{Fig:pdf} and \ref{Fig:radpdf} show the
resulting PDFs for the proper motions, parallaxes and radial
velocities of the Galactic orbits yielding separations $< 10$~pc. Also
indicated are the distributions for a closest approach of $< 5$ pc,
but we find that the shape of the distribution is not altered
significantly. As can be seen, the PDFs for the proper motion of
B2021+51 are not significantly different from our input PDFs. The PDF
for the parallax of B2021+51 indicates a slight preference for a
higher value but this is less than a $0.5\sigma$ shift. The orbital
solutions that yield close proximity of the two pulsars do not require
extreme values for the true proper motion and parallax. For the radial
velocity of B2021+51 we find that the distribution of the close
Galactic orbits peaks around $\pm 100$~\kms\ while higher velocities
are unlikely.  In the case of B2020+28 the PDF of $\mud$ is similar to
the initial Gaussian PDF; however, the distribution of $\mua$ shows a
tendency for more negative values and no solutions are found for $\mua
> -3.5$ mas~yr$^{-1}$. Interestingly, the distribution of the parallax
of B2020+28 is more confined around the observed value than the
observational uncertainties suggest. We can constrain the radial
velocity of B2020+28 to be positive. However, a wide range of positive
velocities can result in a close encounter between the two
pulsars. The B2020+28 radial velocity distribution peaks around 450
\kms. Thus, the astrometric data for both pulsars are completely
consistent with a common origin from a disrupted binary.

We do not find strong correlations between most of the astrometric
parameters. The only obvious correlation we find is between $\pi_1$
and $v_{{\rm r}_2}$, with a larger distance to B2020+28 resulting in a
more negative velocity of B2021+51. A modest correlation also exists
between $\mu_{\alpha_1}$ and $\pi_1$ and as a result between
$\mu_{\alpha_1}$ and $v_{{\rm r}_2}$. Future reduction in the
astrometric uncertainties will allow us to determine a much tighter
relation between $v_{{\rm r}_1}$ and $v_{{\rm r}_2}$.

Fig.~\ref{Fig:closepdf} shows the distributions of the time of closest
approach and the distance to the Sun at that time for the simulations
that result in a minimum pulsar separation $< 10$~pc. We
conclude that the binary disrupted approximately 1.9 Myr ago at
approximately 1.9 kpc distance from the Sun.

\section{The origin of B2020+28 and B2021+51}
\label{sec4}

\subsection{Birth in the Cygnus Superbubble}

\label{cygnus}

 The CSB is a region with intense radio emission surrounded by soft
X-ray emission \citep{Cash80}. The CSB is approximately 2 Myr
old. \citet{UFR+01} estimate that approximately 200 SN have occurred
in the CSB within its lifetime, while $\sim1000$ SN are needed to
power the X-ray emission surrounding the CSB. In fact, Uyaniker et
al.~argue that the CSB is not a single structure at all, but merely
the result from projection effects of several X-ray emitting regions
along the line of sight.

 As shown in Fig.~\ref{Fig:cygnuspath}, the CSB region hosts several
OB associations at distances between 0.7 and 2.5 kpc. The ages of the
OB associations range between 3 and 13 Myr.  The distance and location
of the pulsar pair at its birth site agrees best with the Cyg OB 2
association at $l=80.1^\circ$, $b=0.9^\circ$ and estimated distance
$D=1.44 - 2.10$~kpc, assuming the OB association is at rest with respect
to the Sun. Taking a distance of 1.7 kpc \citep{T-D91}, we find that the
separation between Cygnus OB 2 and the pulsar pair peaks at
approximately 200 pc when the pulsars are at their closest approach,
well within the estimated distance interval to OB 2. Cygnus OB 2 is
between $3-5$~Myr old \citep{T-D91, UFR+01}. It also contains several
of the most massive stars ($>60 M_\odot$) in the Galaxy \citep{ABC81,
MT91}. A common origin of the two pulsars in this association
$\sim2~$Myr ago would require a progenitor binary with two extremely
high mass stars ($>40 M_\odot$) with main-sequence lifetimes less than
3 Myr. Such high mass stars are, however, more likely to form black
holes or disrupt completely when the SN occurs \citep{WW86}. However,
a modest velocity of a few times 10 \kms\ of the OB associations,
could bring some of the older ones in the Cygnus region close to the
pulsar pair. Such velocities are quite reasonable, as the peculiar
velocities of nearby OB associations are shown to be $\sim 0-30$~\kms\
\citep{dZ99}. Additionally, if the pulsars originated from a binary
pair disrupted after the second SN, the first SN can already have
imparted a velocity of several tens of \kms. If for instance, the
second SN took place 0.5~Myr after the first, this would correspond to
a displacement on the sky of almost a degree at the distance of the
Cygnus OB associations.  Unfortunately, in the absence of accurate
velocity information for the Cygnus OB associations, it is impossible
to identify the parent association with certainty.

 We find that two other pulsars in our sample also cross the area of
the CSB for acceptable radial velocities and epochs. Both B1857$-$26
and B2016+28 passed in the vicinity of the CSB and its OB associations
approximately 5~Myr ago. The spin-down ages are 48 and 60~Myr for
B1857$-$26 and B2016+28 respectively. However, their {$\dot{P}$} and
magnetic fields $B$ are somewhat lower than typical values (${\dot{P}} =
2.0$ and $1.5 \times 10^{-16}$~s~s$^{-1}$; $B = 0.35$ and $0.30 \times
10^{-12}$ Gauss for B1857$-$26 and B2016+28), indicating that they
went through a period of recycling. If so, the differences between
their chronological ages and spin-down ages could be large. Therefore,
these pulsars could also have been born in the CSB if they originate
from binaries that went through an episode of accretion which was
interrupted by the second SN.

\subsection{Possible birth scenarios}

 There are several possible scenarios for the common origin of
B2020+28 and B2021+51. If they originate from the same binary, the masses of
the progenitor stars would likely have been within a few percent of
each other to produce two pulsars which have not been recycled and
whose properties and spin-down age are so similar. \citet{BB98} find
that 50\% of all binaries have stars massive enough to undergo SN. Of
these, it is estimated that 2\% have masses similar enough that the
two SN occur within $1$~Myr of each other \citep{B95}.  The binary
could then have been disrupted after the first SN explosion, with the
second SN occurring while the first NS and its companion were already
separated. Alternatively, the binary could have survived the first SN
only to be disrupted by a second SN a short time afterward.
Another possibility is that the pulsars were not part of a binary and
our identification is the result of a chance alignment, as discussed
earlier (\S~\ref{chance}). However, even if the pulsars were not part
of a binary, the likelihood analysis in \S~\ref{chance} indicates that
they both originate from the same OB association in the Cygnus region,
within several pc of each other.

\begin{figure}[t]
\epsscale{0.9} 
\plotone{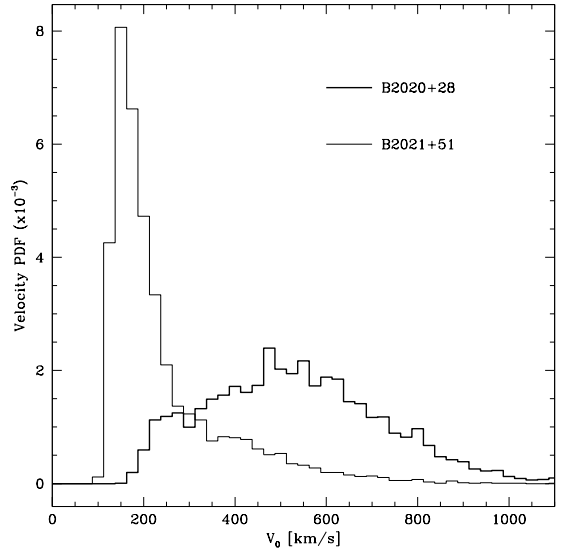}
\caption[Pulsar birth velocity]{ The birth velocity distribution of B2020+28
(thick) and B2021+51 (thin) for the Galactic orbit solutions yielding
separations of $< 10$~pc. The birth velocities have been corrected for
Galactic rotation.}
\label{Fig:vbirth}
\end{figure}

Any birth scenario must explain the pulsar velocities at birth, as
well as the angle between the pulsar velocity vectors. As shown in
Fig.~\ref{Fig:vbirth}, we find that the velocity of B2021+51 at the
time of closest approach peaks strongly at $\approx 150$ \kms. The
velocity of B2020+28 is less well determined, but peaks at
$\approx 500$ \kms.  We find that the distribution of the direction
angle, which is the angle between the birth velocity vectors of the
pulsars, peaks strongly around 160$^\circ$.

\subsubsection{A binary disrupted after the first SN explosion}

\citet{BB98} estimate that approximately 60\% of the binaries
are disrupted after the first SN explosion when the kick velocity
distribution of \citet{CC98} is used in the analysis.
The high space velocities of B2020+28 and B2021+51 should then have
been imparted in the first SN explosion, when the progenitor star of
the second NS was already well into its red-giant phase. 
According to the analysis by \citet{TT98}, a kick velocity
higher than $600$~\kms\ would be needed to impart a velocity similar
to that of B2021+51; however, the first NS would then be expected to
have a velocity of around 700 \kms. Since the velocity of B2020+28 is
not very well constrained, we cannot rule out this scenario on the
basis of the velocities alone. A somewhat lower kick velocity in the
first SN could also explain the high velocities if the second SN (in
this case most likely B2021+51) imparted an additional kick. However,
as we are able to trace the two pulsars to within a few pc of each
other, the second kick must have been in the direction of motion of
the runaway pulsar, unless the second SN occurred within a few
thousand years when the progenitor star had traveled only a few
parsec.

\subsubsection{A binary disrupted after the second SN explosion}

\begin{figure}[tf]
\plotone{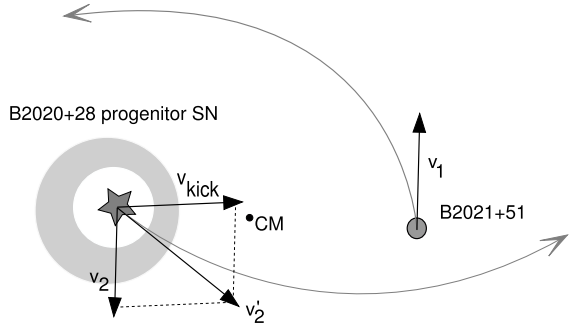}
\caption[binary sketch]{ Schematic representation of a binary
disrupted after the second asymmetric SN explosion (indicated by the
starred symbol with the grey annulus) that can produce the observed
pulsar pair. The high mass progenitor of the second pulsar (B2020+28)
and the NS (B2021+51) have initial circular velocities $v_1$ and $v_2$
respectively, around the center of mass (CM). The kick velocity
$v_{\rm kick}$ imparted in the second SN is in the direction of the NS
companion and the newly created NS moves closely past its companion,
resulting in a slingshot effect that contributes to the observed
pulsar velocities and their resulting orbits (grey curves).}
\label{Fig:binary}
\end{figure}

If the binary survived the first SN explosion, spiral-in and
mass-transfer can occur, and the companion star can evolve into a much
lighter Helium star with a lifetime of the order of 1~Myr. However,
since the pulsars do not exhibit any characteristics of recycling,
this is highly unlikely. We thus expect the binary separation to have
been large enough so that the two stars evolved independently.

\citet{TT98} show that for a system with a large binary separation and
a high-mass companion, a SN kick of only $\sim 200$ \kms\ can already
produce the implied birth velocities that we find for B2020+28 and
B2021+51. Using a simple binary model with an initial circular binary
orbit we have attempted to reproduce the pulsar velocities and the
angle between the birth velocity vectors.  If the disruption of the
binary was caused only by the (instantaneous) mass-loss in a symmetric
SN, the angle between the pulsars' birth velocity vectors cannot be
larger than $90^\circ$ \citep{GGO70}. However, when we include an
asymmetric kick, larger angles can be produced. We find that the model
can reproduce the required velocities and direction angle when the
companion mass $\sim 9$~M$_\odot$, the binary separation is $2-4$~AU
and the kick velocity is $\approx 200$~\kms\ with an angle $\approx
70^\circ$ from the direction of motion of the exploding star toward
the companion NS. As the newly created NS moves closely past the first
NS, both experience a slingshot effect which accelerates each to the
desired velocities. In this simple analysis, we expect B2021+51 to be
created in the first SN explosion while B2020+28 was created in the
explosion that disrupted the binary system. This scenario is
illustrated in Fig.~\ref{Fig:binary}.

 If the initial binary orbit was elliptical, the orbital position as
well as the kick angle must be included in the analysis. It is quite
likely that the binary orbit was no longer circular after the first SN
had failed to disrupt the system. However, since the circular
binary scenario can already easily produce the required velocities and
direction angle, an elliptical binary will most likely be able to do
the same. Thus, the scenario in which our pulsar pair originates from
a binary that was disrupted in the second SN explosion seems the most
likely.

\section{Conclusions}
\label{concl}

 After examining the available sample of pulsars that have precision
astrometric data, we have found that the pulsars B2020+28 and B2021+51
likely share their origin in a disrupted binary even though they are
presently separated by $\sim 23^\circ$ on the plane of the sky. The
Galactic orbits of the pulsars, traced back through a 3 component
St\"ackel potential that satisfies the current Galactic parameters,
are found to cross approximately 1.9 Myr ago at 1.9 kpc distance from
the Sun in the region known as the Cygnus superbubble. Simulations of
pulsar samples, and an analysis of the likelihood functions, show that
it is unlikely that our pulsar pair identification is the result from
a chance alignment.

The most likely scenario for the creation of the runaway pulsar pair
is the disruption of a binary system in the second SN explosion. As no
signs of accretion are present, the progenitor stars must have had
similar masses and have been in a wide binary orbit. The second SN
must have followed the first by only a short time ($<0.5$~Myr).  The
velocities and direction angle of the pulsars at birth can be
explained if the second (asymmetric) SN imparted a kick velocity of
$\approx 200$~\kms. The velocity analysis indicates that B2020+28 was
created in the second SN.

The system was disrupted $\approx~1.9$~Myr ago, which must correspond
to the true age of at least one of the pulsars. The spin-down ages of
both pulsars are very similar: B2020+28 has a spin-down age of
$2.88$~Myr, while B2021+51 has a spin-down age of $2.75$~Myr. As the
relationship of spin-down age to true age is largely uncertain, the similarity
between our derived age and the spin-down ages is remarkable.  An age
of 1.9 Myr implies that the initial period of B2020+28 was 201~ms,
assuming a constant braking index of $n=3$. If the SN that created
B2021+51 was the source of the binary disruption, the initial pulsar
period will have been 295~ms, also assuming a constant $n=3$. In both
cases, a constant braking index of $n>4$ can be ruled out.

The most likely three dimensional space velocities of the pulsars at
birth are $\approx 150$~\kms\ for B2021+51 and $\approx 500$~\kms\ for
B2020+28. Recent analyses of pulsar velocities have found that a two
component birth velocity distribution seems to best fit the current
data \citep{CC98, ACC02}. Using the sample of pulsars with accurate
astrometric data, Chatterjee et al. (2004) find that the pulsar
velocities are best described by a 2-component birth velocity
distribution with characteristic velocities of 86~\kms\ (70\%) and
292~\kms\ (30\%). The probability of finding pulsars with a three
dimensional space velocity $\geq 500$~\kms\ ranges from $\sim 10\%$
using the Chatterjee et al. (2004) distribution, which has probable
selection effects against the highest velocity pulsars, to $\sim 45\%$
for the \citet{ACC02} distribution, which was based on a much larger
sample if pulsars, but used a different methodology that was based
largely on DM-distances using the \citet{TC93} model. The birth
velocities determined for B2020+28 and B2021+51 then indicate that
B2021+51, likely created in the first SN explosion, is part of the
lower velocity population, while B2020+28, created in the SN explosion
that disrupted the binary system, is part of the high velocity
component.

The current astrometric data are completely consistent with the common
binary origin of the two pulsars and an improvement of astrometric
precision is unlikely to invalidate our conclusions. However, as the
accuracies of the proper motion and parallax measurements are
increased, we will be able to more precisely determine the current
radial velocities of the pulsars. This will lead to a significant
improvement on the determination of the birth velocities and as a
result, we will be able to obtain better constraints for the kick
velocities of the SN explosions as well as other characteristics of
the progenitor binary system.

As the sample of pulsars with accurate proper motions and parallaxes
increases, the identification of additional related pulsar pairs will
allow us to put constraints on kick velocities and the amount of
asymmetry in the SN explosions. Unfortunately, due to pulsar beaming,
we can only detect approximately one out of five pulsars. This ratio
is somewhat better for young pulsars, which, as seen in
\S~\ref{chance}, are also the pulsars for which a positive pair
identification is the most likely. For a compete sample of nearby,
young pulsars with precision astrometric data, we expect to be able
to identify related pulsars for at least $10\%$ of them.

\acknowledgements

The National Radio Astronomy Observatory is a facility of the National
Science Foundation (NSF) operated under cooperative agreement by
Associated Universities, Inc.  This work at Cornell was supported in
part by NSF grants AST 9819931 and AST 0206036.

\end{document}